\documentclass[10pt,conference]{IEEEtran}

\usepackage{amssymb}
\usepackage{multirow}
\usepackage{xspace}
\usepackage{amsmath}
\usepackage{adjustbox}
\usepackage{wrapfig}
\usepackage{microtype}
\usepackage{graphics}
\usepackage{balance}
\usepackage{float}
\usepackage{caption}
\usepackage{subcaption}
\usepackage{listings}
\usepackage{graphicx}
\usepackage{grffile}
\usepackage{makecell}
\usepackage{multicol}
\usepackage[table]{xcolor}
\usepackage{colortbl}
\usepackage{framed}
\usepackage{amsthm}
\usepackage{enumitem}
\usepackage{hyperref}
\usepackage{array}
\usepackage{booktabs}
\usepackage[framemethod=TikZ]{mdframed}
\usepackage{cite}
\usepackage{tcolorbox}
\definecolor{bg}{gray}{0.95}
\definecolor{darkred}{RGB}{139, 0, 0} 
\colorlet{shadecolor}{gray!20}
\definecolor{Gray}{gray}{0.8}

\newcommand{\llm}{LLM\xspace}
\newcommand{\llms}{LLMs\xspace}

\newcommand{\misconfig}{misconfiguration\xspace}

\newcommand{\misconfigs}{misconfigurations\xspace}

\newcommand{\config}{configuration\xspace}

\newcommand{\confvalid}{configuration validation\xspace}
\newcommand{\confvaliders}{configuration validators\xspace}
\newcommand{\valid}{validation\xspace}

\newcommand{\ourframework}{Ciri\xspace}
\newcommand{\fw}{Ciri\xspace}

\newcommand{\misconf}{{Misconfig}\xspace}
\newcommand{\validconfig}{{ValidConfig}\xspace}

\newcommand{\filelevel}{file-level\xspace}
\newcommand{\paramlevel}{parameter-level\xspace}
\newcommand{\filelevelcap}{File-Level\xspace}
\newcommand{\paramlevelcap}{Parameter-Level\xspace}
\newcommand{\filelevelabb}{F.L.\xspace}
\newcommand{\paramlevelabb}{P.L.\xspace}

\usepackage{color}
\definecolor{gray}{gray}{0.5}
\definecolor{LightGray}{gray}{0.75}
\definecolor{VeryLightGray}{gray}{0.90}
\definecolor{VeryLightLightGray}{gray}{0.95}

\newcommand{\precision}{precision\xspace}

\newcommand{\recall}{recall\xspace}

\newcommand{\fonescore}{F1-score\xspace}
\newcommand{\ghit}{G-hits\xspace}

\newcommand{\hcommon}{HCommon\xspace}
\newcommand{\hbase}{HBase\xspace}

\newcommand{\para}[1]{\smallskip\noindent {\bf #1} }

\newcommand{\Comment}[1]{}
\newcommand{\Space}[1]{}

\newcommand{\distance}{1pt}
\newtheoremstyle{findingstyle}
  {0pt}   
  {0pt}   
  {\itshape}  
  {0pt}       
  {\bfseries} 
  {.}         
  {5pt plus 1pt minus 1pt} 
  {}          

\theoremstyle{findingstyle}
\newtheorem{findinner}{\textbf{Finding}}

\newenvironment{find}
  {\begin{shaded}\begin{findinner}}
  {\end{findinner}\end{shaded}}

\newcommand{\finding}[1]{
  \begin{find}
    #1
  \end{find}
}

\newcounter{subfinding}[findinner]
\renewcommand\thesubfinding{\thefindinner.\arabic{subfinding}}

\newcommand{\implication}[1]{}

\newcommand{\gptturbo}{GPT-3.5-Turbo\xspace}
\newcommand{\gptfour}{GPT-4\xspace}

\lstdefinelanguage{diff}[]{Java}{
  numbers=left,
  basicstyle={\footnotesize\ttfamily},
  numbersep=6pt,
  breaklines=true,
  xleftmargin=.03\textwidth,
  escapeinside={(*@}{@*)},
  showstringspaces=false,
  tabsize=2,
  frame=none,
}

\begin{document}

\title{\huge Configuration Validation with Large Language Models}

\author{\IEEEauthorblockN{Xinyu Lian, Yinfang Chen, Runxiang Cheng, Jie Huang, Parth Thakkar$^{\dagger}$, Minjia Zhang, Tianyin Xu \vspace{5pt}}
\IEEEauthorblockA{University of Illinois at Urbana-Champaign $^{\dagger}$Meta Platforms, Inc.}\vspace{-15pt}}

\maketitle

\begin{abstract}

Misconfigurations are major causes of software failures.
Existing practices rely on developer-written 
    rules or test cases to validate configurations,
    which are expensive.
Machine learning (ML) for configuration validation
    is considered a promising direction, but has been facing challenges such as 
    the need of large-scale field data and system-specific models. 
Recent advances in Large Language Models (LLMs) show 
    promise in addressing some of the 
    long-lasting limitations of ML-based configuration validation.
We present a first analysis on the feasibility and effectiveness of using 
    \llms for \confvalid. 
We empirically evaluate \llms as \confvaliders by
developing a generic \llm-based configuration \valid framework, named \fw. 
\fw employs effective prompt engineering with few-shot learning based on both 
    valid configuration and misconfiguration data.
\fw checks outputs from LLMs when producing results,
    addressing hallucination and nondeterminism of LLMs.
We evaluate \fw's validation effectiveness on eight popular LLMs 
    using configuration data of ten widely deployed open-source systems.
Our analysis (1) confirms the potential of using LLMs for configuration validation,
    (2) explores design space of LLM-based validators like Ciri, 
    and (3) reveals open challenges such as ineffectiveness in detecting certain types 
    of misconfigurations and biases towards popular configuration parameters.
\end{abstract}

\section{Introduction}
\label{sec:intro}

Modern software systems undertake
  hundreds to thousands of configuration changes on a daily basis~\cite{tang:sosp:15,maurer:15,barroso:2018,sherman:05,mehta:2020,beyer:2018,sayagh18}.
For example, at Meta/\nobreakdash Facebook, thousands of configuration file ``diffs'' are committed
  daily, outpacing the frequency of code changes~\cite{tang:sosp:15,maurer:15}.
Other systems such as at Google and Microsoft also frequently
    deploy configuration changes~\cite{mehta:2020,barroso:2018,beyer:2018}.
Such velocity of configuration changes inevitably leads to misconfigurations.
Today, misconfigurations are among the dominating causes of production incidents~\cite{maurer:15,gunawi:16,yin:sosp:11,xu:15,oppenheimer:03,kendrick:12,rabkin:13,sayagh18,barroso:2018}.

To detect misconfigurations,
    today's configuration management systems employ the ``configuration-as-code'' paradigm
    and enforce continuous configuration validation,
    ranging from static validation, to configuration testing, and to manual review and approval~\cite{tang:sosp:15}.
The configuration is first checked by validation code (aka {\it validators}) 
    based on predefined correctness rules~\cite{tang:sosp:15,raab:17,baset:middleware:2017,huang:15,Leuschner:2017,liao:18,NadiTSE2015,zhang:oopsla:21};
    in practice, validators are written by engineers~\cite{tang:sosp:15,raab:17,baset:middleware:2017}.
After passing validators, configuration changes are then tested with 
    code to check program behavior~\cite{sun:osdi:20,xu:19}.
Lastly, configuration changes are reviewed like source-code changes.

    
The aforementioned pipeline either relies on manual inspection 
    to spot misconfigurations in the configuration file diffs,
    or requires significant engineering efforts to implement and maintain validators or test cases.
However, these efforts are known to be costly and incomprehensive.
For example, despite that mature projects all include extensive configuration validators,
    recent work~\cite{xu:sosp:13,xu:osdi:16,li:18,keller:dsn:08,li:issta:21,wang:ase:23,wangt:icse:23} repeatedly shows 
    that existing validators are insufficient.
The reasons are twofold. 
First, with large-scale systems exposing hundreds to thousands of configuration parameters~\cite{xu:fse:15},
    implementing validators for every parameter becomes a significant overhead.
Recent studies~\cite{xu:sosp:13,tang:sosp:15} report that many parameters 
    are uncovered by existing validators, even in mature software projects. 
Second, it is non-trivial to validate a parameter, which could have many different correctness properties,
    such as type, range, semantic meaning, dependencies with other parameters, etc.;
    encoding all of them into validators is laborious and error-prone,
    not to mention the maintenance cost~\cite{zhang:icse:14,zhang:icse:21}.


Using machine learning (ML) and natural language processing (NLP)
    to detect misconfigurations has been considered a promising approach to addressing the 
    above challenges.
Compared to manually written static validators,
    ML or NLP-based approaches are automatic, easy to scale to a large number of parameters, 
    and applicable to different projects.
Several ML/NLP-based misconfiguration detection techniques were proposed~\cite{Bhagwan:21,zhang:asplos:14,wang:osdi:04,le:06,xiang:atc:20,palatin:kdd:06,potharaju:vldb:15,wang:lisa:03}.
The key idea is to first learn correctness rules 
    from field configuration data~\cite{kiciman:04,zhang:asplos:14,wang:osdi:04,le:06,palatin:kdd:06,wang:lisa:03,Bhagwan:21,santo:16,santo:17}
    or from documents~\cite{xiang:atc:20,potharaju:vldb:15},
    and then use the learned rules to detect misconfigurations
    in new configuration files.
ML/NLP-based approaches have achieved good success.
For example, Microsoft adopted PeerPressure~\cite{wang:osdi:04,huang:ndss:05} as a part of Microsoft Product Support Services (PSS) toolkits.
It collects configuration data in Windows Registry from a large number of Windows users to learn statistical golden states of system configurations.

However, ML/NLP-based misconfiguration detection is also significantly limited.
First, the need for large volumes of system-specific configuration data makes it hard to apply those techniques outside
    corporations that collect user configurations (e.g., Windows Registry~\cite{wang:lisa:03}) 
    or maintain a knowledge base~\cite{potharaju:vldb:15}.
For example, in cloud systems where the {\it same} set of configurations is maintained by a small 
    DevOps team~\cite{tang:sosp:15,sherman:05}, there is often no enough information for learning~\cite{xu:osdi:16}.
Moreover, prior ML/NLP-based detection techniques all target specific projects,
    and rely on predefined features~\cite{palatin:kdd:06}, templates~\cite{zhang:asplos:14}, or models~\cite{potharaju:vldb:15},
    making them hard to generalize.

Recent advances on Large Language Models (LLMs), such as GPT~\cite{chatgpt} and Codex~\cite{codex},
    show promises to address some of the long-lasting limitations 
    of traditional ML/NLP-based misconfiguration detection techniques.
Specifically, LLMs are trained on massive amounts of public data,
    including configuration data---configuration files 
    in software repositories, configuration documents, 
    knowledge-based articles, Q\&A websites for resolving configuration issues, etc. 
Hence, LLMs encode extensive knowledge of both {\it common} and {\it project-specific} configuration.
Such knowledge can be utilized for configuration validation without the need for manual rule engineering.
Furthermore, LLMs show the capability of {\it generalization} and {\it reasoning}~\cite{wei:23,huang:acl:23}
    and can potentially ``understand'' configuration semantics.
For example, they can not only understand that values of a port must be in the range of [0, 65535],
        but also reason that a specific configuration value represents a port (e.g., based on the name and description)
        and thus has to be within the range.

Certainly, \llms have limitations. They 
    are known for hallucination and non-determinism~\cite{bang:arxiv:23,zhang:arxiv:23}. 
Additionally, LLMs have limited input context, which can pose challenges when encoding extensive contexts like 
    configuration file and related code.
Moreover, they are reported to be biased to popular content in the training dataset. 
Fortunately, active efforts~\cite{anthropic100k,nakano2022webgpt,wang:arxiv:23,liu2023trustworthy,manakul2023selfcheckgpt} 
    are made to address these limitations.

In this paper, we present a first analysis on the feasibility and effectiveness of using 
    \llms such as GPT and Claude for \confvalid.    
As a first step, we empirically evaluate \llms in the role of \confvaliders,
    without additional fine-tuning or code generation. 
We focus on basic misconfigurations (those violating explicit correctness constraints) which 
    are common misconfigurations encountered
    in the field~\cite{yin:sosp:11}.
We do not target environment-specific misconfigurations or
    bugs triggered by configuration (\S\ref{sec:discussion}).


To do so, we develop \fw, an LLM-empowered configuration validation framework.
\fw takes a configuration file or a file diff
    as the input;
it outputs detected \misconfigs along with the reasons 
    that explain them.
\fw integrates different 
    LLMs such as GPT-4, Claude-3, and CodeLlama.
\fw devises effective prompt engineering with few-shot learning based on 
    existing configuration data.
\fw also validates the outputs of LLMs to generate validation results,
    coping with the hallucination and non-determinism of LLMs.
A key design principle of \fw is separation of policy and mechanism.
\fw can serve as an open framework for experimenting with different 
    models, prompt engineering, training datasets, 
    and validation methods.
    

We study \fw's validation effectiveness using eight popular LLMs including remote models (GPT-4, GPT-3.5, Claude-3-Opus, and Claude-3-Sonnet),
    and locally housed models (CodeLlama-7B/13B/34B and DeepSeek).
We evaluate
    ten widely deployed open-source systems with diverse types.
Our study confirms the potential of using LLMs for configuration validation, e.g., \fw with 
    Claude-3-Opus detects 45 out of 51 real-world misconfigurations, 
    outperforming recent configuration validation techniques.
Our study also helps understand the design space of LLM-based validators like \fw, 
    especially in terms of prompt engineering with few-shot learning and voting.
We find that using configuration data as shots 
    can enhance validation effectiveness.
Specifically, few-shot learning using both valid configuration 
    and misconfiguration data 
    achieves the highest effectiveness.
Our results also reveal open challenges:
\fw struggles with certain types of misconfigurations such as dependency violations 
    and version-specific misconfigurations.
It is also biased to the popularity of configuration parameters,
    causing both false positives and false negatives.

In summary, this paper makes the following contributions:
\begin{itemize}[leftmargin=*]
    \item A new direction of configuration validation using pre-trained large language models (LLMs);
    \item \fw, an LLM-empowered configuration validation framework and an open platform for configuration research;
    \item An empirical analysis on the effectiveness of LLM-based configuration validation,
        and its design space;
    \item We have released \fw and other research artifacts at \url{https://github.com/ciri4conf/ciri}.
\end{itemize}
\begin{figure*}[t!]
    \centering
    \includegraphics[width=\linewidth]{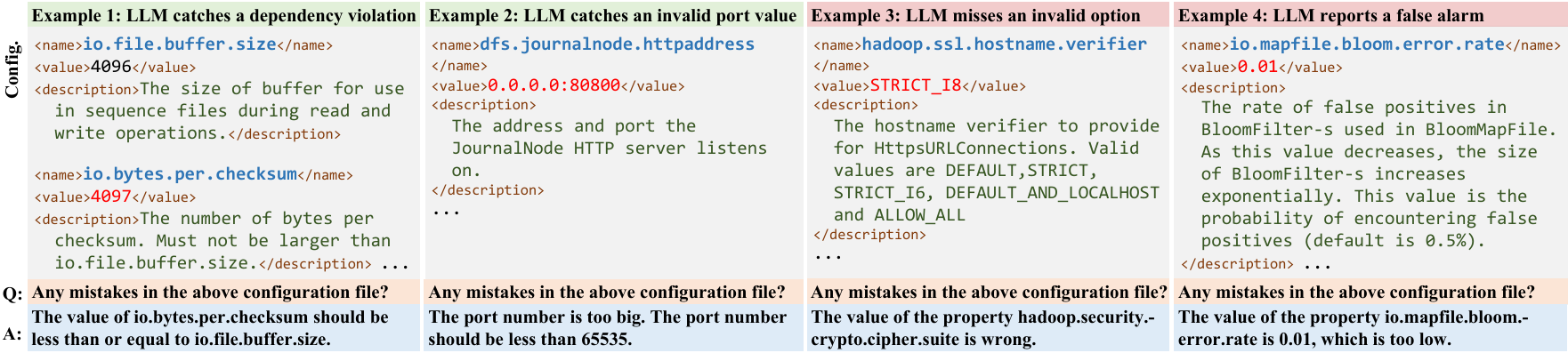}
    \caption{Example 1 and 2 show the \llm correctly catches and reasons the misconfigurations.
        Example 3 and 4 show the \llm misses the misconfiguration or reports a valid configuration as erroneous.}
    \label{fig:example}
    \vspace{10pt}
\end{figure*}

\section{Exploratory Examples}
\label{sec:background:motivation}

We explore using \llms to validate configuration out of the box.
We show that vanilla \llms can detect \misconfigs. 
However, they are prone to both false negatives and false positives that require careful handling.
Figure~\ref{fig:example} presents four examples, two of which the \llm successfully detects misconfigurations, 
    and two of which the \llm misses the misconfiguration or reports a false alarm. 
These examples were generated using the GPT-3.5-Turbo \llm~\cite{ziegler:arxiv:19}.

\para{Detecting violation of configuration dependency.}
Validating dependencies between configuration parameters has been a challenging task in highly-configurable systems~\cite{chen:fse:20,xu:15}.
\llms can infer relations between entities from text at the level of human experts~\cite{brown:arxiv:20},
    which allows \llms to infer dependencies between parameters in a given configuration file 
    based on their names and descriptions.
Figure~\ref{fig:example} (Example 1) presents a case where values of two dependent parameters were changed (i.e., ``{\it buffer.size}'' and ``{\it bytes.per.\allowbreak checksum})''.
After understanding the value relationship dependency between these two parameters, the model determines that the change in ``{\it bytes.per.checksum}'' has violated the enforced dependency, and provides the correct reason for the misconfiguration.

\para{Detecting violation with domain knowledge.}
A state-of-the-art \llm is trained on a massive amount of textual data and possesses basic knowledge across a wide range of professional domains.
An \llm thus could be capable of understanding the definition of a configuration parameter and reasoning with its semantics.
When the \llm encounters a configuration parameter such as IP address, permissions, and masks, it invokes the domain knowledge specific to the properties of those parameters.
Figure~\ref{fig:example} (Example 2) presents a case where an \texttt{HTTP} address has been misconfigured to a semantically invalid value.
The model detects the misconfiguration, reasons that its value is out of range, and further suggests a potential fix.

\para{Missed misconfiguration and false alarm.}
\llms as configuration validators are not without errors.
Examples 3 and 4 in Figure~\ref{fig:example} show two cases where the \llm makes mistakes.

In Example 3, the configuration file has provided a description of the changed parameter ``{\it hostname.verifier}''
and explicitly listed the valid value options of the parameter. 
However, the model fails to realize that the parameter is misconfigured to an invalid, non-existent option ({STRICT\_I8}). 
In Example 4, the description suggests that the parameter ``{\it bloom.error.rate}''
    ranges from 0 to 100 (percentage), whereas the actual scale is 0 to 1 (fraction). 
This inconsistency supposedly confuses the model making it mark 0.01 (a valid value) as invalid. 

Both examples show that directly using off-the-shelf \llms as configuration validators 
would result in false negatives and false positives.
The incorrect validation results can be attributed to hallucination~\cite{gpt4}. 
A simple explanation is that \llms are exposed to potentially contradictory data during training, 
    which causes confusion to the model at the inference time.


\section{Ciri: A LLM-empowered Configuration Validation Framework}
\label{sec:sv}

We develop Ciri, an LLM-empowered configuration validation framework.
\ourframework takes a configuration file or a file diff
    as the input, and
outputs a list of detected \misconfigs along with the reasons 
    to explain the misconfigurations.
\ourframework supports 
    different LLMs such as GPT, Claude, CodeLlama, and DeepSeek~\cite{gpt4,roziere2024codellama}.\footnote{Adding 
    a new \llm in \ourframework takes
    a few lines to add the query APIs.}

\begin{figure}[tb!]
    \centering
    \vspace{-17.5pt}
    \includegraphics[width=\linewidth]{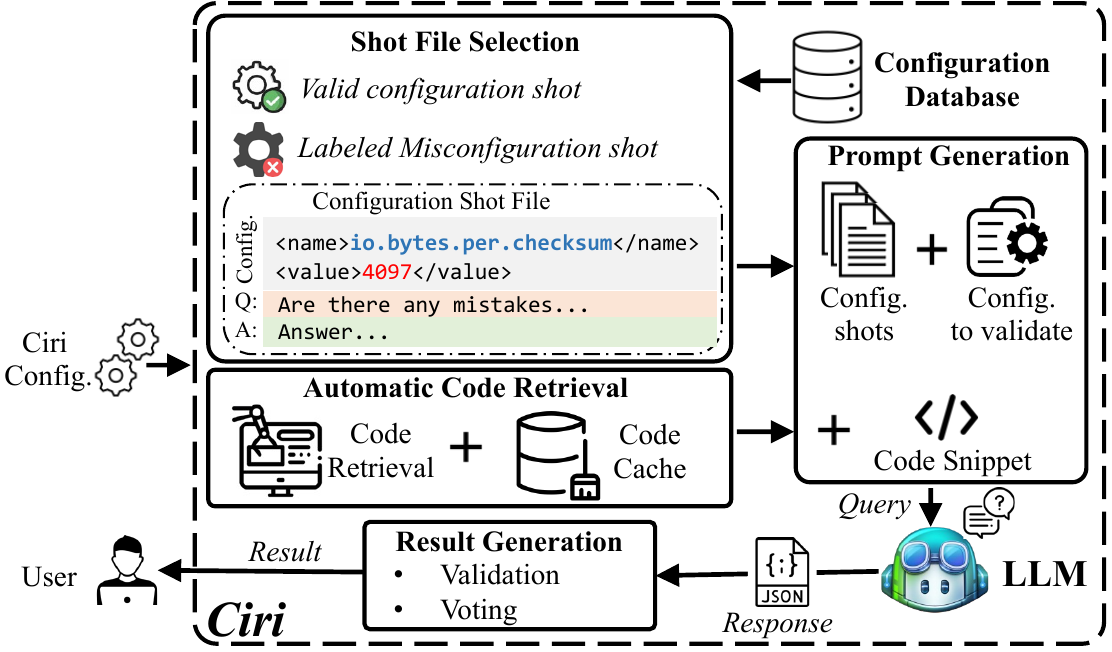}
    \vspace{-15pt}
    \caption{System overview of \ourframework.}
    \label{fig:ciri}
\end{figure}

Figure~\ref{fig:ciri} gives an overview of \ourframework.
\ourframework turns a configuration validation request into a prompt to the \llms (\S\ref{sec:sv:pp}).
The prompt includes
    (1) the target configuration file or diff,
    (2) a few examples (aka \textit{shots}) to demonstrate the task of configuration validation,
    (3) code snippets automatically extracted from codebase,
    and (4) directive question and metadata.
To generate shots, \ourframework uses its database that contains labeled 
    configuration data, including both valid configurations and misconfigurations.
\ourframework sends the same query to the LLMs multiple times
    and aggregates responses into the final validation result (\S\ref{sec:sv:result}).

\ourframework applies to any software project, even if
    it has no labeled configuration data of that project in its database.
\ourframework exhibits
transferability (using data from one project and applying it to others),
the ability to transfer configuration-related knowledge across projects 
when using configurations from different projects as shots (Finding~\ref{finding:transfer}). 
\ourframework's configuration validation effectiveness can also be further improved 
by generating quality shots 
(Finding~\ref{finding:zeroshot}) 
and code snippets (Finding~\ref{finding:code-retrieval}).

\subsection{Prompt Engineering}
\label{sec:sv:pp}


\subsubsection{Prompt structure} 
\ourframework generates a prompt that includes four elements:
    a) the content of input \config file or file diff,
    b) the shots as valid configurations or \misconfigs 
    with questions and ground truth responses for few-shot learning,
    c) code snippets automatically extracted from available codebase,
    and d) a directive question for \llm to respond in formatted output.
Figure~\ref{fig:prompt} shows an illustrative example of the prompt generated by \ourframework.
It contains $N$ shots, the content of to-be-validated configuration file, and the code snippet enclosed within $\langle$Usage$\rangle$ followed by the directive question.

\fw phrases the prompting question as
    \textit{``Are there any mistakes in the above configuration for {[PROJECT]} version {[VERSION]}? 
    Respond in a JSON format similar to the following: ...''}. 
The {[PROJECT]} and {[VERSION]} are required inputs of \ourframework because the validity of configuration can change by project and project version~\cite{zhang:icse:21,zhang:icse:14}.
This prompt format enforces the \llm to respond in a unified JSON format for result aggregation (\S\ref{sec:sv:result}). 
However, responses from \llms sometimes may still deviate from the anticipated format~\cite{bang:arxiv:23,zhang:arxiv:23}. 
In such cases, \ourframework retries a new query to the \llm.


\subsubsection{Few-shot learning}
\begin{figure}[tb!]
    \centering
    \includegraphics[width=\linewidth]{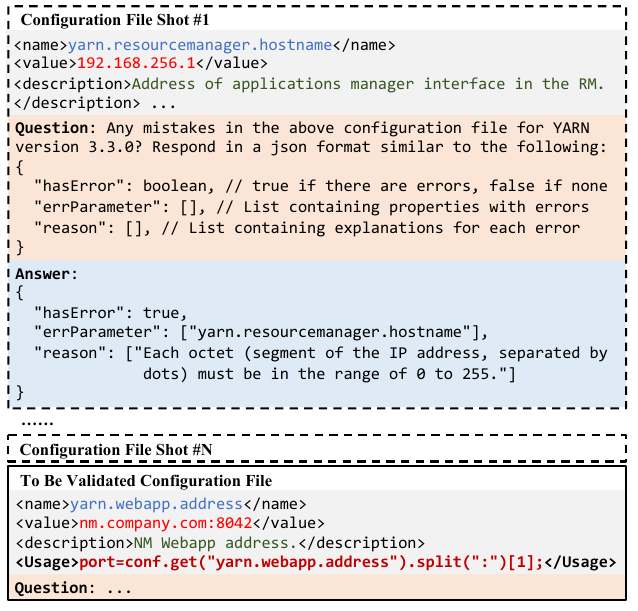}
    \caption{An example prompt generated by Ciri.}
    \label{fig:prompt}
    \vspace{3pt}
\end{figure}

\ourframework leverages the \llm's ability to learn from examples at inference time (aka few-shot learning) 
    to improve configuration validation effectiveness. 
To do so, \ourframework simply inserts shots at the beginning of each prompt. 
Each shot contains a configuration snippet, the prompting question, and its corresponding ground truth. 
Figure~\ref{fig:prompt} shows an example, where 
    there are $N$ shots. 
``Configuration File Shot \#1'' is the first shot, 
in which the parameter ``{yarn.resourcemanager.hostname}'' is misconfigured. 
This shot also contains the prompting question (orange box) and the ground truth (blue box). 


\subsubsection{Shot generation}
\ourframework maintains a database of labeled valid configurations and misconfigurations for generating valid configuration shots ({\it \validconfig}) 
    and misconfiguration shots ({\it \misconf}).
A \validconfig shot specifies a set of configuration parameters and their valid values. 
A valid value of a parameter can be its default value, or other valid values used in practice.
A \misconf shot specifies a set of parameters and their values,
    where only one of the parameter values is invalid.

For a given configuration of a specific project, 
    \ourframework by default generates shots using configuration data of the same project.
If \ourframework's database does not contain configuration data for the target project,
    \ourframework will use data from other projects to generate shots. 
As shown in Finding~\ref{finding:transfer}, 
    \llms possess transferrable knowledge in configuration across different projects.

\fw supports multiple methods for selecting data to generate shots,
    including randomized selection,
    category-based selection, 
    and similarity-based selection (selecting data from configuration with the highest 
    cosine similarity).
We did not observe major differences when using different selection methods.
So, \ourframework uses randomized selection by default.

\subsubsection{Augmenting with code}
Recent work shows that retrieval-augmented generation (RAG) 
    can enhance \llms by incorporating additional information~\cite{lewis2021retrievalaugmented,guu2020realm}.
In the context of configuration, 
    each configuration parameter has corresponding program context, such as 
%
    data type, semantics, and usage. 
In Figure~\ref{fig:prompt}, the code snippet enclosed within $\langle$Usage$\rangle$ is automatically extracted from 
    the source code,
    which uncovers semantics that the parameter value is expected to include a ``:'' symbol, 
    with the split segment representing a port.

\fw uses a simple but effective code retrieval strategy: 
To retrieve the most effective code snippet, \fw employs the following strategy:
    a) searching codebase with parameter names and retrieving relevant snippets;
    b) prioritizing code over comments and documents;
    c) selecting the longest snippet if multiple options are available, as longer snippets tend to be more comprehensive with details,
    and d) deduplicating retrieved snippets.
The retrieved code snippet will be stored in a cache for efficiency.
The code retrieval strategy is effective in improving the effectiveness of configuration validation (Finding~\ref{finding:code-retrieval}).

\subsubsection{Addressing token limits}
\llms limit input size per query by the number of input tokens. 
For example, the token limits for \gptturbo are 16,385. 
To navigate these constraints, \ourframework compresses the prompt if its size exceeds the limit.
\ourframework first tries to put the target configuration 
    and the directive question in the prompt, then 
    maximizes three \misconf shots with one \validconfig shot (Finding~\ref{finding:zeroshot}) to fit into the remaining space.
If the configuration cannot fit into the token limit, 
    \ourframework transforms it into a more compact format, 
    e.g., transforming an {XML} file into {INI} format. 
If the compressed input still cannot fit, \ourframework aborts and returns errors.
In practice, configuration files and diffs are small~\cite{tang:sosp:15,xu:fse:15} and can easily fit existing limits. 
For example, prior study inspects configuration files collected from Docker, where
    each file contains 1 to 18 parameters, with eight on average~\cite{openctest}. 
For very large configurations, \ourframework can split them into multiple snippets and validate them separately. 

\subsection{Result Generation}
\label{sec:sv:result}
The JSON response from \llms contains three primary fields:
a) ``{\it hasError}'': a boolean value indicating whether misconfigurations are detected,
b) ``{\it errParameter}'': a list of misconfigured parameters,
and c) ``{\it reason}'': a list of explanations of the detected misconfiguration, 
    corresponding to ``{\it errParameter}''.

\subsubsection{Validation against hallucination}
We employ a few rules to address the hallucination of \llms.
For example, if {\it hasError} is false, 
    both {\it errParameter} and \texttt{\it reason} must be empty.
Similarly, if {\it hasError} returns true, 
    {\it errParameter} and {\it reason}
    must be non-empty with the same size.
The answer to ``{\it errParameter}'' must not contain repeated values.
If a response fails these rules, \ourframework retries until 
    the \llm returns a valid response.


\subsubsection{Voting against inconsistency}
\label{sec:ciri:clustering}
\llms can produce inconsistent outputs in conversation~\cite{adiwardana:arxiv:20}, 
    explanation~\cite{camburu:arxiv:19},
    and knowledge extraction~\cite{elazar:tacl:21}.
To mitigate inconsistency, 
    \ourframework uses a multi-query strategy---querying the \llm multiple times using the same prompt
    and aggregating responses towards a result 
    that is both representative of the model's understanding and more consistent than a single query.
\ourframework uses a frequency-based voting strategy: 
    the output that recurs most often among the responses is selected as the final output~\cite{wang:arxiv:23}.

Note that the ``{\it reason}'' field is not considered during voting 
    due to the diverse nature of the response.
After voting, \fw collects reasons from all responses 
    associated with the selected {\it errParameter}. 
The reason field is important as it provides users with insights 
    into the misconfiguration, which is different from the traditional ML approaches
    that only provide a binary answer with a confidence score.
\ourframework clusters the {reasons} based on TF-IDF similarity~\cite{wikitfidf2024}, 
    and picks a {reason} from the dominant cluster. 
We find that this mechanism is robust to hallucination--- hallucinated {reasons} were 
    often filtered out as they tended to be very different from each other. 

\subsection{\ourframework Configuration}
\label{sec:system_variables}

\begin{table}[b!]
    \vspace{5pt}
    \centering
    \setlength{\tabcolsep}{2.5pt}
    \scriptsize
    \caption{configuration of \fw and its default values.}
    \vspace{-5pt}
    \label{tab:system_variables}
    \begin{tabular}{llc}
    \toprule
    \textbf{Parameter} & \textbf{Description} & \textbf{Default Value} \\
    \midrule
    Model & Backend \llm. Also allows users to add other \llms. & \gptfour \\
    Temperature & Tradeoff between creativity and determinism. & 0.2 \\
    \# Shots & The number of shots included in a prompt. & Dynamic \\
    \# Queries & The number of queries with the same prompt. & 3 \\
    \bottomrule
    \end{tabular}
\end{table}

\begin{table*}[t!]
  \setlength{\tabcolsep}{2.1pt}
  \renewcommand{\arraystretch}{1.15} 
  \footnotesize
  \centering
  \caption{Misconfiguration generation (we use generation rules from prior work~\cite{li:issta:21,li:18,sun:osdi:20,keller:dsn:08},
    which reflects real-world misconfiguraions). 
  ``Subcategory'' lists rules to generate different misconfiguraions for the same configuration parameter.} 
  \vspace{-5pt}
  \label{tab:misconfig_gen}
  \begin{tabular}{llll}
  \toprule
  {\bf Category}  & {\bf Subcategory}  & {\bf Specification}   & {\bf Generation Rules} \\
  \midrule \\[-3.5ex]
  \rowcolor{VeryLightGray} \cellcolor{white} \multirow{7}{*}{Syntax}   &   & Value set = \{Integer, Float, Long...\}   & Generate a value that does not belong to the value set \\
  \rowcolor{VeryLightGray} \cellcolor{white}& \multirow{-2}{*}{Data type}  & Numbers with units  & Generate an invalid unit (e.g., "nounit") \\
  & Path  & \verb|^(\/[^\/ ]*)+\/?$|  & Generate a value that violates the pattern (e.g., \verb|/hello//world|) \\
  & \cellcolor{VeryLightGray}URL   &\cellcolor{VeryLightGray} \verb|[a-z]+://.*|  & \cellcolor{VeryLightGray}Generate a value that violates the pattern (e.g., \verb|file///|) \\
  & IP address  & \verb|[\d]{1,3}(.[\d]{1,3}){3}|   & Generate a value that violates the pattern (e.g., \verb|127.x0.0.1|) \\
  & \cellcolor{VeryLightGray}Port  & \cellcolor{VeryLightGray}Data type, value set = \{Octet\}  & \cellcolor{VeryLightGray}Generate a value that does not belong to the value set \\
  & Permission  & Data type, value set = \{Octet\}  & Generate a value that does not belong to the value set \\
\cline{1-4}
\addlinespace[0.2ex]
  \multirow{7}{*}{Range}  & \cellcolor{VeryLightGray}Basic numeric   & \cellcolor{VeryLightGray}Valid Range constrainted by data type   & \cellcolor{VeryLightGray}Generate values outside the valid range (e.g., Integer.MAX\_VALUE+1) \\
  & Bool  & Options, value set = \{true, false\}  & Generate a value that does not belong to the value set \\
  & \cellcolor{VeryLightGray}Enum  & \cellcolor{VeryLightGray}Options, value set = \{``enum1'', ``enum2'', ...\}  & \cellcolor{VeryLightGray}Generate a value that doesn't belong to set \\
  & IP address  &  Range for each octet = [0, 255]  & Generate a value outside the valid range (e.g., \verb|256.123.45.6|) \\
  & \cellcolor{VeryLightGray}Port  &  \cellcolor{VeryLightGray}Range = [0, 65535]   & \cellcolor{VeryLightGray}Generate a value outside the valid range \\
  & Permission  &  Range = [000, 777]   & Generate a value outside the valid range \\
\cline{1-4}
\addlinespace[0.2ex]
  \multirow{2}{*}{Dependency} & \cellcolor{VeryLightGray}Control   & \cellcolor{VeryLightGray}$(P_1, V, \Diamond) \mapsto P_2$, $\Diamond \in \{>, \ge, =, \neq, < ,\leq \}$  & \cellcolor{VeryLightGray}Generate invalid control condition $(P_1, V, \lnot\Diamond)$ \\
  & Value Relationship  & $(P_1, P_2, \Diamond)$, $\Diamond \in \{>, \ge, =, \neq, < ,\leq \}$  & Generate invalid value relationship $(P_1, P_2, \lnot\Diamond)$ \\
  \cline{1-4}  
\addlinespace[0.2ex]
  Version  & \cellcolor{VeryLightGray}Parameter change  &  \cellcolor{VeryLightGray}$(V_1, Pset_1) \mapsto (V_2, Pset_2)$, $Pset_1 \neq Pset_2$  & \cellcolor{VeryLightGray}Generate a removed parameter in $V_2$ or use an added paraemter in $V_1$ \\ [-0.4ex]
  \bottomrule
  \end{tabular}
  \vspace{8pt}
\end{table*}

\ourframework is customizable, with a key principle of separating policy and mechanism.
Users can customize \ourframework via its own configurations.
Table~\ref{tab:system_variables} shows several important \ourframework configurations 
    and default values.
The default values are chosen by pilot studies using a subset of our dataset (\S\ref{sec:meth}).

\section{Benchmarks and Metrics}
\label{sec:meth}

 
Our study evaluates ten mature and widely deployed open-source projects:     
    Alluxio, Django, Etcd, HBase, Hadoop Common, HDFS, PostgreSQL, Redis, YARN, ZooKeeper,
    which are implemented in a variety of programming languages (Java, Python, Go, and C).
    They also use different configuration formats (XML and INI) with a large number of configuration parameters.
Table~\ref{tab:dataset} lists the version ({SHA}) and the number of parameters at that version.

We evaluate \fw on the aforementioned projects with eight LLMs: 
    GPT-4-Turbo, GPT-3.5-Turbo, Claude-3-Opus, Claude-3-Sonnet, CodeLlama-7B/13B/34B, and DeepSeek-6.7B, which differ in model sizes and capabilities. 
All of these models have also been trained with a large amount of code data, 
    where prior work has demonstrated their promising capability in handling a number of software engineering tasks~\cite{xia:icse:23,thakur:arxiv:23,chen:arxiv:23}.

\subsection{Configuration Dataset}
Our study uses two types of datasets: 
    real-world misconfiguration datasets 
    and synthesized misconfiguration datasets.

\subsubsection{Real-world misconfiguration}
\label{sec:meth:rwdata}
To our knowledge, the Ctest dataset~\cite{openctest} is the only public dataset of real-world misconfigurations;
    it is used by prior configuration research~\cite{Bhagwan:21,cheng:issta:21,sun:osdi:20,wang:icse:23}.
The dataset contains 64 real-world configuration-induced failures of five open-source projects,
    among which 51 are misconfigurations, and 13 are bugs.
We discuss the results of \fw on real-world misconfigurations in Finding~\ref{finding:real_world}.

\begin{table}[t!]
\centering
\footnotesize
\setlength{\tabcolsep}{2pt}
\caption{Evaluated projects and the configuration datasets (ValidConfig and Misconfig)
    for shot pool and evaluation.}
\vspace{-5pt}
\label{tab:dataset}
\begin{tabular}{lccccccc}
\toprule
\multirow{2}{*}{\textbf{Project}} & {\textbf{Version}} & \multirow{2}{*}{\ \textbf{\# Params}} \ & \multicolumn{2}{c}{\textbf{\validconfig}} & & \multicolumn{2}{c}{\textbf{\misconf}} \\ 
\cline{4-5} 
\cline{7-8}
 & \bf (SHA) &  &  \textbf{\# Shot} & \textbf{\# Eval} & & \textbf{\# Shot}& \textbf{\# Eval} \\
\midrule
\href{https://github.com/Alluxio/alluxio}{Alluxio} & {76569bc} & 494 &  13 & 54 & & 13 & 54\\
\href{https://github.com/django/django}{Django} & {67d0c46} & 140 &  6 & 18 & & 6 & 18\\
\href{https://github.com/etcd-io/etcd}{Etcd} & {946a5a6} & 41 &  8 & 32 & & 8 & 32\\
\href{https://github.com/apache/hbase}{HBase} & {0fc18a9}  & 221 &  12 & 50 & & 12 & 50 \\ 
\href{https://github.com/apache/hadoop/tree/trunk/hadoop-common-project}{HCommon} & {aa96f18} & 395 & 16 & 64 & & 16 & 64 \\ 
\href{https://github.com/apache/hadoop/tree/trunk/hadoop-hdfs-project}{HDFS}  & {aa96f18}  & 566 & 16 & 64& & 16 & 64\\
\href{https://github.com/postgres/postgres}{PostgreSQL}  & {29be998}  & 315 & 8 & 31& & 8 & 31\\
\href{https://github.com/redis/redis}{Redis}  & {d375595}  & 94 & 12 & 44&& 12 & 44\\
\href{https://github.com/apache/hadoop/tree/trunk/hadoop-yarn-project}{YARN}  & {aa96f18}  & 525& 10&40 & & 10 & 40\\
\href{https://github.com/apache/zookeeper}{ZooKeeper} & {e3704b3} & 32& 8& 32& & 8 & 32\\
\bottomrule
\end{tabular}
\vspace{5pt}
\end{table}

\subsubsection{Synthesized misconfiguration}
Since real-world configuration dataset (\S\ref{sec:meth:rwdata}) is too small,
to systematically evaluate configuration validation effectiveness,
    we create new synthetic datasets for each evaluated project.
First, we collect default configuration values from the default configuration file 
    of each project, and real-world configuration files from the Ctest dataset (collected from Docker images~\cite{openctest})
    for those projects included in Ctest.
We then generate misconfigurations of different types.
The generation rules are from prior studies on misconfigurations~\cite{xu:sosp:13,li:18,keller:dsn:08,li:issta:21},
    which violates the constraints of configuration parameters (Table~\ref{tab:misconfig_gen}).
Notably, prior studies show that the generation rules can cover 96.5\% of 1,582 parameters across four projects~\cite{li:18}.

For each project, we build two distinct configuration sets.
First, we build a configuration dataset with no misconfiguration (denoted as \validconfig)
    to measure true negatives and false positives (Table~\ref{tab:confusion}). 
We also build a configuration dataset (denoted as \misconf) 
    in which each configuration file has one misconfiguration,
    to measure true positives and false negatives (Table~\ref{tab:confusion}).
Note that a misconfiguration can be a dependency violation between multiple parameter values.

\addtocounter{table}{+1} 
\begin{table*}
\caption{F1-score, precision, and recall of \fw evaluated on ten projects 
    with eight LLMs as configuration validators.}
\vspace{-5pt}
    \setlength{\tabcolsep}{2.25pt}
    \renewcommand{\arraystretch}{1.15} 
    \scriptsize
    \centering
    \begin{tabular}{l|cccccccccc|c|ccccccccccc|cc|c|ccc|c|cc}
        \toprule
        \multirow{3}{*}{\bf Models}  & \multicolumn{23}{c}{\bf F1-score} &&& \multicolumn{2}{c}{\bf Precision} &&&\multicolumn{2}{c}{\bf Recall} \\
         & \multicolumn{11}{c}{\bf \filelevelcap (\filelevelabb)} && \multicolumn{11}{c}{\bf \paramlevelcap (\paramlevelabb)} &&&  {\bf \filelevelabb} & {\bf \paramlevelabb} &&& {\bf \filelevelabb} & {\bf \paramlevelabb}\\
         & AL. & DJ. & ET. & HB. & HC. & HD. & PO. & RD. & YA. & ZK.& \textbf{Avg} & & AL. & DJ. & ET. & HB. & HC. & HD. & PO. & RD. & YA. & ZK. & \textbf{Avg} &&& \textbf{Avg} & \textbf{Avg} &&& \textbf{Avg} & \textbf{Avg}\\
         \midrule
         GPT-4-Turbo&0.69&0.86&0.67&0.73&0.75&0.70&0.72&0.75&0.74&0.70&0.73&&0.52&0.82&0.59&0.49&0.51&0.53&0.43&0.57&0.62&0.53&0.56&&&0.62&0.44&&&0.89&0.81\\
         GPT-3.5-Turbo&0.68&0.71&0.73&0.77&0.78&0.68&0.65&0.71&0.72&0.74&0.72&&0.48&0.55&0.60&0.55&0.58&0.55&0.36&0.54&0.61&0.66&0.55&&&0.62&0.43&&&0.89&0.77\\
         Claude-3-Opus&0.71&0.81&0.70&0.70&0.79&0.74&0.75&0.82&0.77&0.78&0.76&&0.51&0.62&0.53&0.54&0.65&0.60&0.53&0.62&0.60&0.60&0.58&&&0.65&0.45&&&0.91&0.83\\
         Claude-3-Sonnet&0.74&0.77&0.79&0.80&0.81&0.76&0.82&0.84&0.79&0.78&\cellcolor{Gray}0.79&&0.53&0.65&0.69&0.73&0.69&0.73&0.53&0.64&0.71&0.59&\cellcolor{Gray}0.65&&&0.75&0.57&&&0.85&0.79\\
         CodeLlama-34B&0.69&0.87&0.80&0.64&0.70&0.67&0.79&0.78&0.66&0.83&0.74&&0.61&0.85&0.76&0.35&0.45&0.35&0.59&0.65&0.46&0.79&0.59&&&0.65&0.51&&&0.89&0.70\\
         CodeLlama-13B&0.71&0.67&0.67&0.71&0.66&0.70&0.71&0.76&0.69&0.77&0.70&&0.54&0.59&0.61&0.48&0.37&0.37&0.50&0.69&0.51&0.73&0.54&&&0.61&0.45&&&0.85&0.68\\
         CodeLlama-7B&0.67&0.80&0.74&0.67&0.67&0.67&0.63&0.73&0.67&0.76&0.70&&0.53&0.67&0.66&0.27&0.28&0.23&0.51&0.68&0.43&0.72&0.50&&&0.56&0.40&&&0.96&0.67\\
         DeepSeek-6.7B&0.72&0.72&0.81&0.52&0.47&0.46&0.70&0.67&0.59&0.84&0.65&&0.58&0.56&0.75&0.48&0.40&0.37&0.44&0.44&0.55&0.73&0.53&&&0.76&0.60&&&0.66&0.55\\
      \bottomrule
      \end{tabular}
      \vspace{5pt}
      \label{tab:overview}
\end{table*}

To create the \misconf data for each project, 
    we first check if its configuration parameters fit any subcategory in Table~\ref{tab:misconfig_gen},
    and, if so, we apply rules from all matched subcategories to generate misconfigurations for that parameter.
For example, an IP-address parameter fits both ``Syntax: IP Address'' and ``Range: IP Address''.
We do so for all parameters in the project.
Then, we randomly sample at most five parameters in each subcategory that has matched parameters,
    and generate invalid value(s) per sampled parameter. 
For each subcategory, we further randomly select one parameter 
    from the five sampled ones. 
We use the selected parameter to create a faulty configuration as a \misconf shot (\S\ref{sec:sv}) for that subcategory
    and add it to the project's shot pool.
For the other four parameters, we use them to create four faulty configurations for that subcategory, and use them for evaluation.
If a subcategory does not have enough parameters for sampling, we use all the parameters for evaluation.
We separate the evaluation set and shot pool to follow the practice that the training set does not overlap with the testing set~\cite{brown:arxiv:20}.
We create the \validconfig dataset for each project using the aforementioned methodology for 
    the \misconf dataset, except that we generate valid values.

Table~\ref{tab:dataset} shows the size for both the \validconfig and \misconf datasets for each project. 
Note that our datasets cover 72\%--100\% of the entire parameter set of each project.

\subsection{Metrics}
\label{sec:meth:metrics}
We evaluate Ciri's effectiveness at both configuration \textit{file} and \textit{parameter} levels: 
(1) at the file level, we check if Ciri can determine if a configuration file contains misconfigurations;
(2) at the parameter level, we check if Ciri can determine if each parameter in the configuration file is valid or not. 
Table~\ref{tab:confusion} describes our confusion matrix.
We compute the precision \textit{(TP/(TP+FP))}, recall \textit{(TP/(TP+FN))}, and F1-score at both file and parameter levels. 
If not specified, we default to macro averaging since each project is regarded equally.
We prioritize \paramlevel effectiveness for fine-grained measurements and 
    discuss \paramlevel metrics by default in the evaluation.

\addtocounter{table}{-2} 
\begin{table}[h!]
\vspace{6pt}
\centering
\footnotesize
\setlength{\tabcolsep}{2.3pt}
\caption{Definitions for confusion matrix.\label{tab:confusion}}
\vspace{-5pt}
\begin{tabular}{lll}
\toprule
\multicolumn{1}{l}{\textbf{Level}} & \multicolumn{1}{l}{\textbf{Metric}} & \textbf{Definition} \\
\midrule
\multirow{4}{*}{File} & TP & A misconfigured file correctly identified \\ 
 & FP & A correct file wrongly flagged as misconfigured \\ 
 & TN & A correct file rightly identified as valid \\ 
 & FN & A misconfigured file overlooked or deemed correct \\ 
\midrule
\multirow{4}{*}{Param.} & TP & A misconfigured parameter correctly identified \\ 
 & FP & A correct parameter wrongly flagged as misconfigured \\ 
 & TN & A correct parameter rightly identified as valid \\ 
 & FN & A misconfigured parameter overlooked or deemed correct \\
\bottomrule
\end{tabular}
\vspace{3pt}
\end{table}




\section{Evaluation and Findings}
\label{sec:finding}
We present empirical results on the effectiveness of \llms as configuration validators 
    with \ourframework (\S\ref{sec:eval:overall}). 
We analyze how validation effectiveness changes with regard to 
    design choices of \fw (\S\ref{sec:eval:design}). 
We also present our understanding of when \ourframework produces wrongful results (\S\ref{sec:eval:diff}) 
    and biases (\S\ref{sec:eval:bias}).

\subsection{Effectiveness of Configuration Validation}
\label{sec:eval:overall}

\vspace{-5pt}
\finding{\ourframework shows effectiveness of using state-of-the-art \llms as configuration validators.
    It achieves file- and \paramlevel F1-scores up to 0.79 and 0.65, respectively.\label{finding:overall}}
\vspace{-7.5pt}

\noindent
\ourframework exhibits remarkable capability in \config validation.
Table~\ref{tab:overview} shows the \fonescore, \precision, and \recall for each project using \llms 
    with three \misconf and one \validconfig shots (Finding~\ref{finding:zeroshot}).
The results show that Ciri not only can effectively identify configuration files with \misconfig
    (with an average \fonescore of 0.72 across 8 LLMs), 
    but also pinpoint misconfigured parameters with explanations
    (with an average \fonescore of 0.56 across 8 LLMs).
Certainly, the \paramlevel \fonescore{}s are about 15\% 
    lower than \filelevel \fonescore{}s, i.e., 
    pinpointing fine-grained misconfigured parameters is a more challenging task for \llms compared to 
    classifying the entire file as a whole.


\finding{\fw detects 45 out of 51 real-world misconfigurations,
    outperforming recent configuration validation techniques, including learning-based~\cite{Bhagwan:21}
    and configuration testing~\cite{sun:osdi:20}.
\label{finding:real_world}}
\vspace{-7.5pt}

\noindent
We conduct experiments to evaluate how \fw compares with existing validation techniques on a real-world dataset.
For this evaluation, we choose the top five LLMs ranked by F1-score at the parameter-level based on the results from Table~\ref{tab:overview}.
The real-world dataset~\cite{openctest} 
    contains 51 misconfigurations in total (\S\ref{sec:meth}),
    among which \fw can detect 33-45 misconfigurations, as shown in Table~\ref{tab:real-world}.
\fw successfully detected 45 using Claude-3-Opus.
The six undetected misconfigurations include three due to parameter dependency violations 
    (discussed further in \S\ref{sec:eval:diff}),
    and the other three are environment-related issues that are beyond Ciri's current capability.
Notably, \fw seldom reports incorrect detection.

We compare Ciri's results with a recent learning-based validation technique, ConfMiner~\cite{Bhagwan:21},
    which was evaluated on the same dataset.
ConfMiner utilizes the file content and commit history to identify patterns in configuration to detect misconfigurations.
ConfMiner can detect 27 out of 51 misconfigurations,
    which is 40\% less than \fw.
Unlike \llms that are trained on extensive text data and can comprehend the context of configurations, 
    ConfMiner relies on regular expressions to identify patterns. 
This approach limits its ability in complex scenarios, such as identifying valid values for enumeration parameters 
    and understanding the relationships between different parameters.

\addtocounter{table}{+1} 
\begin{table}[t!]
    \centering
    \footnotesize
    \setlength{\tabcolsep}{3pt}
    \caption{A comparison of \fw, ConfMiner, and Ctest 
        in detecting real-world misconfigurations. N.R.: ``not reported.''}
    \vspace{-5pt}
    \label{tab:real-world}
    \begin{tabular}{llccl}
    \toprule
    \multirow{2}{*}{\textbf{Technique}} & {\textbf{\# Correct}} & {\textbf{\# Incorrect}} & \multirow{2}{*}{\textbf{\# Missed}} & \multirow{2}{*}{\textbf{Runtime}} \\ 
    & \textbf{Detection} & \textbf{Detection} & & \\
    \midrule
    {Ciri (Claude-3-Opus)} & 45 (88.2\%) & 1 & 5 & 20-60 sec\\
    {Ciri (GPT-4-Turbo)} & 41 (80.4\%) & 2 & 8 & 15-40 sec\\ 
    {Ciri (CodeLlama-34B)} & 39 (76.5\%) & 1 & 11 & 30-70 sec\\
    {Ciri (GPT-3-Turbo)} & 37 (72.5\%) & 3 & 11 & 10-25 sec\\ 
    {Ciri (Claude-3-Sonnet)} & 33 (67.7\%) & 1 & 17 & 10-30 sec\\
    ConfMiner & 27 (52.3\%) & N.R. & N.R. & N.R. \\
    Ctest & 41 (80.4\%) & N.R. & N.R. & 20-230 min\\
    \bottomrule
    \end{tabular}
    \vspace{3pt}
    \end{table}

We also compare Ciri with a recent configuration testing technique, namely Ctest~\cite{sun:osdi:20,cheng:issta:21,wang:icse:23}.
Ctest detected 41 of the real-world misconfigurations without rewriting test code;
    Ciri outperforms Ctest by 8.9\%.
The reasons are twofold.
First, testing relies on adequacy of the test cases.
We find that existing test suites do not always have a high coverage 
    of configuration parameters.
On the other hand, LLMs can validate any parameter. 
Second, LLMs detected ``silent misconfigurations''~\cite{yin:sosp:11} that are not manifested via crashes or captured by assertions 
    (e.g., several injected misconfigurations silently fell back to default values and passed the test; 
    LLMs detected them likely because they violated documented specifications).

Certainly, Ctest can detect a broader range of misconfigurations 
    such as the environment-related issues that \fw cannot.
We do not intend to replace configuration testing with LLMs. 
Instead, our work shows that LLMs can provide much quicker feedback for common types of misconfigurations, 
    so tools like Ciri can be used in an early phase (e.g., configuration authoring) 
    before running expensive configuration testing. 
As shown in Table~\ref{tab:real-world}, 
    for a configuration file in the real-world dataset,
    Ctest takes 20 to 230 minutes to finish~\cite{sun:osdi:20}, 
    while Ciri only takes 10 to 70 seconds.

In summary, our results show that \llms like GPT, Claude-3, and CodeLlama-34B
    can effectively validate configurations and detect misconfigurations
    with a sensibly designed framework like \fw.
\fw can provide prompt feedback,
    complementing other techniques like configuration testing.


\subsection{Impacts of Design Choices}
\label{sec:eval:design}

Ciri plays a critical role in LLMs' effectiveness of configuration validation.
We explore its design choices and impacts.

\vspace{-2pt}
\finding{Using configuration data as shots 
    can effectively improve \llms' effectiveness of configuration validation. 
Shots including both valid configuration and misconfiguration 
    achieve the highest effectiveness. \label{finding:zeroshot} }
\vspace{-7.5pt}

\noindent
Using validation examples as shots 
    can effectively improve the effectiveness of \llms.
 Table~\ref{tab:weekness} shows the results of \llms 
    when the validation query does not include shots. 
 In particular, comparing Table~\ref{tab:weekness} to Table~\ref{tab:overview}, 
    as indicated by the numbered arrows in Table~\ref{tab:weekness},
    the average \fonescore of the \llms has decreased by 0.03--0.54 at the file level,
    and decreased by 0.21--0.47 at the parameter level. 

\begin{table}[ht]
    \setlength{\tabcolsep}{4pt}
    \renewcommand{\arraystretch}{1.15} 
    \footnotesize
    \caption{Effectiveness of \llms without using shots.}
    \vspace{-5pt}
    \centering
    \begin{tabular}{l|cc|cc|cc}
      \toprule
      \multirow{2}{*}{\bf Models}  & \multicolumn{2}{c}{\bf F1-score} & \multicolumn{2}{c}{\bf Precision} &\multicolumn{2}{c}{\bf Recall} \\
       & {\bf \filelevelabb} & {\bf \paramlevelabb} &  {\bf \filelevelabb} & {\bf \paramlevelabb} & {\bf \filelevelabb} & {\bf \paramlevelabb}\\
       \midrule
       GPT-4-Turbo&0.70 (\textcolor{darkred}{0.03$\downarrow$)}&0.34 (\textcolor{darkred}{0.22$\downarrow$)}&0.57&0.23&0.93&0.82\\
       GPT-3.5-Turbo&0.67 (\textcolor{darkred}{0.05$\downarrow$)}&0.20 (\textcolor{darkred}{0.35$\downarrow$)}&0.50&0.12&0.99&0.77\\
       Claude-3-Opus&0.69 (\textcolor{darkred}{0.07$\downarrow$)}&0.37 (\textcolor{darkred}{0.21$\downarrow$)}&0.64&0.28&0.82&0.69\\
       Claude-3-Sonnet&0.67 (\textcolor{darkred}{0.12$\downarrow$)}&0.28 (\textcolor{darkred}{0.37$\downarrow$)}&0.55&0.20&0.89&0.66\\
       CodeLlama-34B&0.66 (\textcolor{darkred}{0.08$\downarrow$)}&0.12 (\textcolor{darkred}{0.47$\downarrow$)}&0.50&0.07&0.96&0.52\\
       CodeLlama-13B&0.59 (\textcolor{darkred}{0.11$\downarrow$)}&0.12 (\textcolor{darkred}{0.42$\downarrow$)}&0.53&0.07&0.76&0.52\\
       CodeLlama-7B&0.65 (\textcolor{darkred}{0.05$\downarrow$)}&0.11 (\textcolor{darkred}{0.39$\downarrow$)}&0.51&0.08&0.91&0.23\\
       DeepSeek-6.7B&0.11 (\textcolor{darkred}{0.54$\downarrow$)}&0.06 (\textcolor{darkred}{0.47$\downarrow$)}&0.99&0.50&0.06&0.04\\
    \bottomrule
    \end{tabular}
      \vspace{12pt}
      \label{tab:weekness}
\end{table}

 

\begin{figure}[t!]
\vspace{-9pt}
\centering
\includegraphics[width=0.9\linewidth]{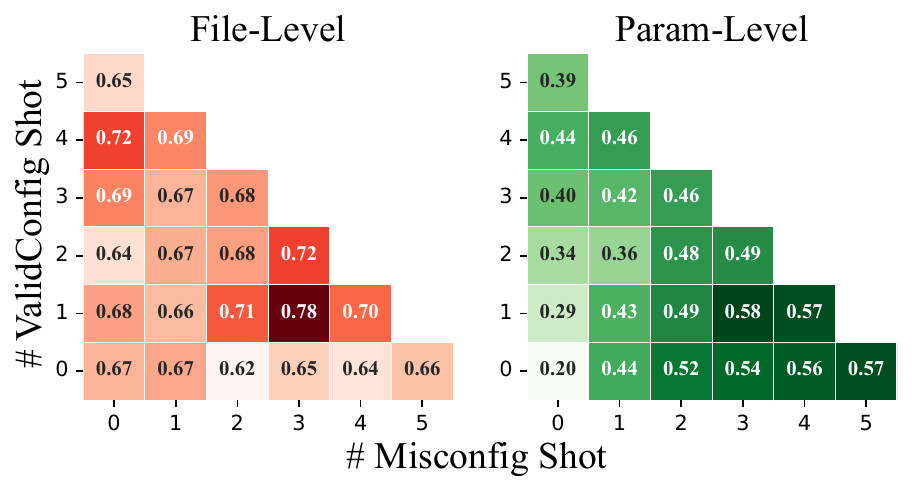}
\vspace{-10pt}
\caption{
F1 scores under different shot combinations.\label{fig:heatmap}}
\vspace{5pt}
\end{figure}

We also study Ciri's effectiveness with different shot combinations.
We evaluate six $N$-shot learning settings, where $N$ ranges from 0 to 5. 
For example, to evaluate \fw with a two-shot setting, 
    three experiments will be performed: 
    (1) two \validconfig shots; 
    (2) one \validconfig shot plus one \misconf shot; 
    (3) two \misconf shots. 
In total, we experiment with 21 shot combinations. 
Due to cost, we only run experiments
    on GPT-3.5-Turbo to HCommon.
We find that only using \validconfig shots leads to a decrease in \precision, 
    while only using \misconf shots reduces \recall.
Clearly, text distribution in the query affects 
    \llms~\cite{min:arxiv:22}.
\llms can be biased:
    if the shots are all misconfigurations,
    \llms will be overly sensitive to the specific patterns in the shots, known as overfitting,
    which causes \llms miss other types of misconfigurations;
    if the shots are all \validconfig, \llms face challenges in accurately identifying incorrect parameters within the file,
    leading to false alarms.
As shown in Figure~\ref{fig:heatmap}, using both \misconf and \validconfig in few-shot learning 
    mitigates the biases and achieves the highest effectiveness, 
    and including three \misconf shots and one \validconfig shot in the prompt 
    achieves the highest \fonescore at both the file and parameter levels.


\vspace{-2.5pt}
\finding{Using configuration data from the same project 
    as shots often leads to high validation F1 score. 
However, even without access to configuration data from the target project, 
    using configuration data from a different project can lead to a improved validation score than zero-shot.
\label{finding:transfer}}
\vspace{-7.5pt}

In situations where configuration data is unavailable (e.g., due to confidentiality),
    we evaluate whether using configuration data from other systems as shots
    can improve configuration validation effectiveness on the target system.
Table~\ref{tab:cross-project shot} shows the results of using data from other projects as shots
    for configuration validation on \hcommon.
By comparing 4-shot HCommon with other columns in Table~\ref{tab:cross-project shot}, 
    we see that using shots from other projects is not as effective as using shots from the target system.
However, the average \fonescore 
    is still higher than zero-shot, indicating that using shots from other projects can improve the effectiveness over zero-shot.
Our observations highlight that \ourframework with \llms can transfer 
    configuration-related knowledge across different projects 
    for effective configuration validation compared to traditional approaches.

\begin{table}[t]
    \setlength{\tabcolsep}{0.5pt}
    \vspace{4pt}
    \footnotesize
    \caption{F-1 score on \hcommon (HC.) using shots from different systems, e.g., HB. refers to using \hbase shots. 4-S and 0-S means using four shots and no shots respectively.}
    \centering
    \begin{tabular}{l|cc||ccc|c||cc||ccc|c}
        \toprule
        \multirow{3}{*}{\bf Models} & \multicolumn{6}{c}{\bf \filelevelcap (\filelevelabb)} & \multicolumn{6}{c}{\bf \paramlevelcap (\paramlevelabb)} \\
        & \multicolumn{2}{c||}{HC.} & \multirow{2}{*}{Dj.} &  \multirow{2}{*}{ET.} & \multirow{2}{*}{HB.} & \multirow{2}{*}{\textbf{Avg}} &  \multicolumn{2}{c||}{HC.} & \multirow{2}{*}{Dj.} &  \multirow{2}{*}{ET.} & \multirow{2}{*}{HB.} & \multirow{2}{*}{\textbf{Avg}}\\
         & 4-S & 0-S &  &   &  & & 4-S & 0-S &  &  &  & \\
         \midrule
        {GPT-3.5-Turbo} & 0.78& \cellcolor{VeryLightGray}0.67 & 0.78&0.68&0.74&\cellcolor{Gray}0.74&0.58 & \cellcolor{VeryLightGray}0.20&0.44&0.42&0.51&\cellcolor{Gray}0.46\\
        {Claude-3-Sonnet} & 0.81 & \cellcolor{VeryLightGray}0.67 & 0.74&0.74&0.78&\cellcolor{Gray}0.75 &0.69 & \cellcolor{VeryLightGray}0.28&0.54&0.59&0.67&\cellcolor{Gray}0.60\\
      \bottomrule
      \end{tabular}
      \label{tab:cross-project shot}
      \vspace{5pt}
\end{table}

\begin{table*}[t]
  \setlength{\tabcolsep}{0.75pt}
  \renewcommand{\arraystretch}{1.15} 
  \fontsize{6pt}{7.2pt}\selectfont
  \caption{Parameter-level \fonescore by \misconfig types from Table~\ref{tab:misconfig_gen}. 
  N.A. means no evaluation samples.}
  \vspace{-5pt}
  \centering
  \begin{tabular}{ll|cccccccccc|c|c|cccccccccc|c|c|cccccccccc|c|c}
      \toprule
      \multirow{2}{*}{\bf Category} & \multirow{2}{*}{\bf Sub-category} & \multicolumn{11}{c|}{\bf GPT-4-Turbo} & & \multicolumn{11}{c|}{\bf Claude-3-Opus} & & \multicolumn{11}{c|}{\bf CodeLlama-34B} \\
      \cline{3-13}
      \cline{15-25}
      \cline{27-37}
      & &    AL. & DJ. & ET. & HB. & HC. & HD. & PO. & RD. & YA. & ZK. & \bf Avg & & AL. & DJ. & ET. & HB. & HC. & HD. & PO. & RD. & YA. & ZK. & \bf Avg & & AL. & DJ. & ET. & HB. & HC. & HD. & PO. & RD. & YA. & ZK. &  \bf Avg \\
      \midrule
      \multirow{7}{*}{\textbf{Syntax}} 
      &Data Type&0.84&1.00&1.00&0.67&0.94&0.89&0.70&0.78&0.89&0.80&\cellcolor{Gray}\textbf{0.85}&&0.67&0.80&1.00&0.80&0.94&1.00&0.80&0.74&1.00&0.80&\cellcolor{Gray}\textbf{0.85}&&0.93&1.00&1.00&0.25&0.67&0.80&1.00&1.00&0.50&1.00&\cellcolor{Gray}\textbf{0.82}\\
      &Path&0.50&1.00&1.00&0.89&0.73&0.73&0.57&0.57&1.00&0.73&\cellcolor{Gray}\textbf{0.77}&&1.00&1.00&0.80&0.89&0.89&1.00&1.00&0.67&0.89&0.73&\cellcolor{Gray}\textbf{0.89}&&1.00&1.00&1.00&0.86&0.50&0.00&1.00&0.57&1.00&1.00&\cellcolor{Gray}\textbf{0.79}\\
      &URL&0.67&0.80&1.00&N.A.&0.80&0.80&N.A.&N.A.&N.A.&N.A.&\cellcolor{Gray}\textbf{0.81}&&1.00&0.67&0.73&N.A.&1.00&0.89&N.A.&N.A.&N.A.&N.A.&\cellcolor{Gray}\textbf{0.86}&&1.00&1.00&0.75&N.A.&1.00&0.60&N.A.&N.A.&N.A.&N.A.&\cellcolor{Gray}\textbf{0.87}\\
      &IP Address&0.70&N.A.&N.A.&0.73&0.94&0.89&N.A.&0.84&0.94&0.76&\cellcolor{Gray}\textbf{0.83}&&0.73&N.A.&N.A.&0.84&0.94&0.84&N.A.&0.80&0.94&0.84&\cellcolor{Gray}\textbf{0.85}&&1.00&N.A.&N.A.&0.82&0.75&0.75&N.A.&1.00&1.00&1.00&\cellcolor{Gray}\textbf{0.90}\\
      &Port&0.74&N.A.&N.A.&0.78&0.94&0.82&N.A.&0.94&N.A.&0.84&\cellcolor{Gray}\textbf{0.84}&&0.70&N.A.&N.A.&0.82&0.82&0.67&N.A.&0.84&N.A.&0.94&\cellcolor{Gray}\textbf{0.80}&&0.94&N.A.&N.A.&0.71&1.00&0.62&N.A.&0.75&N.A.&1.00&\cellcolor{Gray}\textbf{0.84}\\
      &Permission&0.89&N.A.&N.A.&0.80&0.78&1.00&N.A.&N.A.&N.A.&N.A.&\cellcolor{Gray}\textbf{0.87}&&0.89&N.A.&N.A.&0.80&0.82&1.00&N.A.&N.A.&N.A.&N.A.&\cellcolor{Gray}\textbf{0.88}&&0.86&N.A.&N.A.&0.50&0.50&0.40&N.A.&N.A.&N.A.&N.A.&\cellcolor{VeryLightGray}\textbf{0.56}\\
      \cline{1-37}
      \multirow{7}{*}{\textbf{Range}} 
      &Basic Numeric&0.73&1.00&0.75&0.73&0.60&0.67&1.00&0.89&1.00&0.80&\cellcolor{Gray}\textbf{0.82}&&0.67&0.80&0.89&0.46&0.50&0.67&1.00&0.67&0.89&0.80&\cellcolor{Gray}\textbf{0.73}&&0.75&1.00&0.89&0.36&0.00&0.50&0.67&0.75&0.50&1.00&\cellcolor{Gray}\textbf{0.64}\\
      &Bool&1.00&0.75&1.00&0.80&1.00&1.00&N.A.&0.89&1.00&0.80&\cellcolor{Gray}\textbf{0.92}&&1.00&0.55&0.73&0.67&0.89&0.89&N.A.&0.73&1.00&0.80&\cellcolor{Gray}\textbf{0.80}&&1.00&0.86&1.00&0.00&0.67&0.00&N.A.&0.67&0.50&0.75&\cellcolor{Gray}\textbf{0.60}\\
      &Enum&0.36&N.A.&0.86&0.73&0.67&0.89&0.89&0.75&0.80&N.A.&\cellcolor{Gray}\textbf{0.74}&&0.36&N.A.&0.89&1.00&0.86&0.75&0.89&0.89&0.80&N.A.&\cellcolor{Gray}\textbf{0.80}&&0.86&N.A.&1.00&0.75&0.57&0.60&0.75&1.00&0.75&N.A.&\cellcolor{Gray}\textbf{0.78}\\
      &IP Address&0.70&N.A.&N.A.&0.73&0.94&0.89&N.A.&0.84&0.94&0.76&\cellcolor{Gray}\textbf{0.83}&&0.73&N.A.&N.A.&0.84&0.94&0.84&N.A.&0.80&0.94&0.84&\cellcolor{Gray}\textbf{0.85}&&1.00&N.A.&N.A.&0.82&0.75&0.75&N.A.&1.00&1.00&1.00&\cellcolor{Gray}\textbf{0.90}\\
      &Port&0.74&N.A.&N.A.&0.78&0.94&0.82&N.A.&0.94&N.A.&0.84&\cellcolor{Gray}\textbf{0.84}&&0.70&N.A.&N.A.&0.82&0.82&0.67&N.A.&0.84&N.A.&0.94&\cellcolor{Gray}\textbf{0.80}&&0.94&N.A.&N.A.&0.71&1.00&0.62&N.A.&0.75&N.A.&1.00&\cellcolor{Gray}\textbf{0.84}\\
      &Permission&0.89&N.A.&N.A.&0.80&0.78&1.00&N.A.&N.A.&N.A.&N.A.&\cellcolor{Gray}\textbf{0.87}&&0.89&N.A.&N.A.&0.80&0.82&1.00&N.A.&N.A.&N.A.&N.A.&\cellcolor{Gray}\textbf{0.88}&&0.86&N.A.&N.A.&0.50&0.50&0.40&N.A.&N.A.&N.A.&N.A.&\cellcolor{VeryLightGray}\textbf{0.56}\\
      \cline{1-37}
      \multirow{2}{*}{\textbf{Dependency}} 
      &Control&0.50&N.A.&0.00&0.00&0.00&0.25&0.00&0.00&0.00&N.A.&\cellcolor{VeryLightLightGray}\textbf{0.09}&&0.00&N.A.&0.29&0.00&0.00&0.00&0.00&0.00&0.00&N.A.&\cellcolor{VeryLightLightGray}\textbf{0.04}&&0.40&N.A.&0.67&0.29&0.00&0.33&0.00&0.40&0.00&N.A.&\cellcolor{VeryLightLightGray}\textbf{0.26}\\
      &Value Relationship&1.00&N.A.&N.A.&0.75&0.36&0.22&0.40&N.A.&0.50&N.A.&\cellcolor{VeryLightGray}\textbf{0.54}&&0.80&N.A.&N.A.&0.67&0.29&0.25&0.40&N.A.&0.33&N.A.&\cellcolor{VeryLightGray}\textbf{0.46}&&0.00&N.A.&N.A.&0.29&0.44&0.57&0.67&N.A.&0.67&N.A.&\cellcolor{VeryLightGray}\textbf{0.44}\\
      \cline{1-37}
      \textbf{Version} 
      &Parameter Change&0.00&N.A.&0.29&0.44&0.36&0.00&0.29&0.33&0.00&N.A.&\cellcolor{VeryLightLightGray}\textbf{0.21}&&0.00&N.A.&0.75&0.00&0.75&0.00&0.00&0.80&0.57&N.A.&\cellcolor{VeryLightGray}\textbf{0.36}&&0.00&N.A.&0.67&0.00&0.29&0.00&0.00&0.00&0.00&N.A.&\cellcolor{VeryLightLightGray}\textbf{0.12}\\
  \bottomrule
\vspace{5pt}
  \end{tabular}
  \label{tab:result_category}
\end{table*}

\vspace{-3pt}    
\finding{Ciri's code augmentation approach can help \llms to better understand the context of the configuration 
    and improve the validation effectiveness.\label{finding:code-retrieval}}

\begin{figure}[h!]
   \vspace{2pt}
    \centering
        \includegraphics[width=\linewidth]{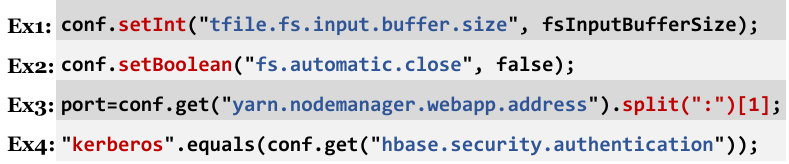}
        \vspace{-16pt}
        \caption{Code snippets retrieved by \ourframework to aid \llms.}
        \label{fig:code-retrieval}
    \vspace{5pt}
\end{figure}

\vspace{-8pt}
\noindent
We compare the results of \gptturbo with and without code augmentation with four shots.
The results show an improvement in F1 scores by 0.03 at both the file and parameter levels.
Figure~\ref{fig:code-retrieval} exemplifies code snippets retrieved from the codebase by Ciri, 
    which could improve \llms' comprehension of the configuration context:
    (1) examples 1 and 2 delineate the parameter types as Integer and Boolean, respectively.
    (2) example 3 highlights that the parameter should include a ``:" symbol, with the latter segment representing a port.
    (3) example 4 shows that ``kerberos" is one valid value for the parameter.

\finding{Code-specialized LLMs, e.g., CodeLlama, 
    exhibit much higher validation scores than generic LLMs, e.g., Llama-2. 
Moreover, further scaling up the code-specialized LLMs leads to a continuous increase 
    in validation scores.\label{finding:model-size}}
\vspace{-8pt}
\noindent
In Table~\ref{tab:overview}, 
    we observe a notable trend within the CodeLlama model family:
the 13B model demonstrates an improvement in \fonescore at the parameter level, 
    by 0.04 over the 7B model;
this trend continues with the 34B model, 
    which exhibits a further 0.05 enhancement in \fonescore over the 13B model.
The observed performance gains can be primarily attributed to the increased capacity for 
    learning and representing complex semantics of configuration values as model size scales.
This involves deep comprehension beyond syntax and range violations
    which are more common in practice~\cite{yin:sosp:11}.

We further evaluated the effectiveness of the Llama-2 model, 
    which is identical to CodeLlama in structure but lacks code-specific training.
The Llama-2-13B is not effective,
    with an average \fonescore of 0.05 at the parameter level.
This result underscores the role of code-specific training, 
    which enhances LLM's comprehension of configuration in the context of code.

\subsection{Limitations and Challenges}
\label{sec:eval:diff}

\vspace{-7.5pt}
\finding{With \ourframework, \llms excel at detecting \misconfigs of syntax and range violations 
    with an average \fonescore of 0.8 across subcategories. 
However, \llms are limited in detecting \misconfigs of dependency and version violations 
with an average \fonescore of 0.3 across subcategories.\label{finding:category}}
\vspace{-7.5pt}
\noindent
Table~\ref{tab:result_category} shows Ciri's validation effectiveness 
    per misconfiguration type.
The \fonescore on detecting misconfigurations 
    of syntax and range violations is consistently above 0.5 across projects,
    and often reaches 0.8. 
However, \fonescore rarely exceeds 0.5 on misconfigurations of dependency and version violations.
Under these two categories, 
    LLMs achieve \fonescore{}s of 0.44--0.54 
    for misconfigurations that violate value relationship, 
    which is higher than the other two subcategories 
    (control and parameter change);
    however, it is still much lower than others.

\begin{figure}[h!]
    \centering
        \includegraphics[width=\linewidth]{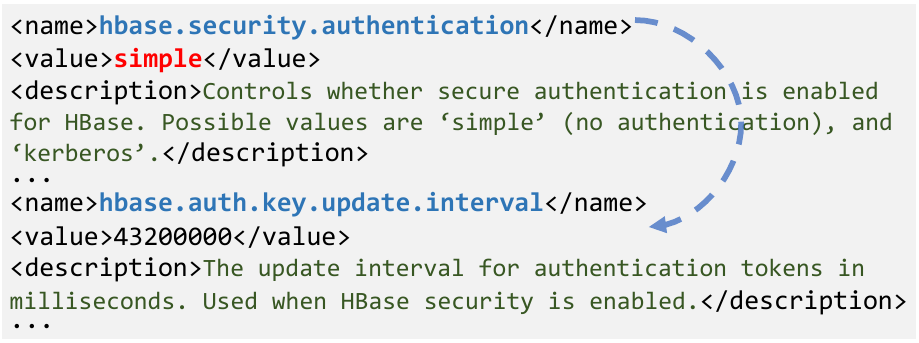}
        \vspace{-15pt}
        \caption{Misconfiguration that violates control dependency that \llms cannot detect. 
        The update interval for authentication is set but the secure authentication is disabled.}
        \label{fig:dependency-misconfig}
\end{figure}

The difference can be attributed to 
    the inherent nature of different types of \misconfigs. 
Misconfigurations of syntax and range violations are more common in practice~\cite{yin:sosp:11}, 
    from which \llms learned extensive knowledge. 
In such a case, domain-specific knowledge from \llms is sufficient to detect these misconfigurations. 
But, misconfiguration of dependency and version violations is often project-specific,
    as exemplified in Figure~\ref{fig:dependency-misconfig}. 
They are tied to detailed history and features of the project, 
    and thus hard to be captured or memorized by \llms 
    if the \llm is not fine-tuned on project-specific data.
This discrepancy between \misconfig types 
    exposes existing \llm's limitation. 

\begin{figure}[t]
    \vspace{-8pt}
    \includegraphics[width=\linewidth]{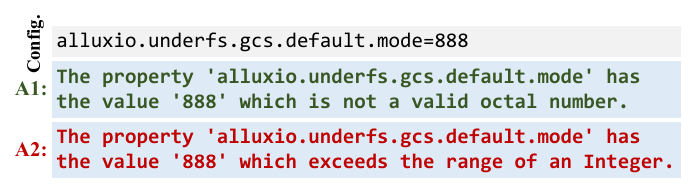}
    \vspace{-15pt}
    \caption{Correct and incorrect reasons returned by \llms.}
    \label{fig:wrong-reason}
\end{figure}

\vspace{-7pt}
\finding{Among the detected misconfigurations, 
    \llms correctly explained reasons for 93.9\% of the misconfigurations;
    meanwhile, 6.1\% of the reasons are misleading.\label{finding:reason}}
\vspace{-7.5pt}
\noindent
When an \llm detects a misconfiguration, 
    \ourframework also asks the \llm to explain the reason 
    to aid debugging and fixing the misconfiguration (\S\ref{sec:sv:pp}).  
To evaluate these explanations, 
    we randomly select one answer in which the misconfiguration is correctly detected 
    per $\langle$subcategory, project, LLM$\rangle$ tuple, and collect a total of 740 answers (resulting from 2,220 queries). 
Upon careful manual review, we determined that 93.9\% of the reasons given by the LLMs are clear 
    and explain the misconfigurations. 
3.1\% of the answers contain a mix of correct and incorrect reasons across queries. 
Ciri filters out incorrect reasons using the voting mechanism (\S\ref{sec:ciri:clustering}) 
    as correct reasons are dominating. 
Figure~\ref{fig:wrong-reason} presents an example of mixed reasons, 
    with the second reason being an instance of hallucination.

In short, with frameworks like \fw, LLMs can effectively validate configurations 
    for syntax or range violations;
    but are less effective for the configurations that involve 
    dependencies between parameters and software versions, showing the challenges for LLMs to reason 
    about interactions
    between parameters and between configuration and code~\cite{meinicke:16}.
To address those misconfigurations, 
    one can re-train or fine-tune LLMs with data related to dependency and versions. 

\subsection{Biases}
\label{sec:eval:bias}

\vspace{-8pt}
\finding{\llms are biased to popular parameters: 
Ciri is more effective in detecting misconfigurations of popular parameters,
    but also reports more false alarms on them.}
\vspace{-7.5pt}

\noindent
To measure the popularity of a configuration parameter, 
    we count the number of exact-match search results 
    returned by Google when searching the parameter name 
    and call it {\it \ghit}.

\begin{figure}[t!]
    \centering
    \includegraphics[width=\linewidth]{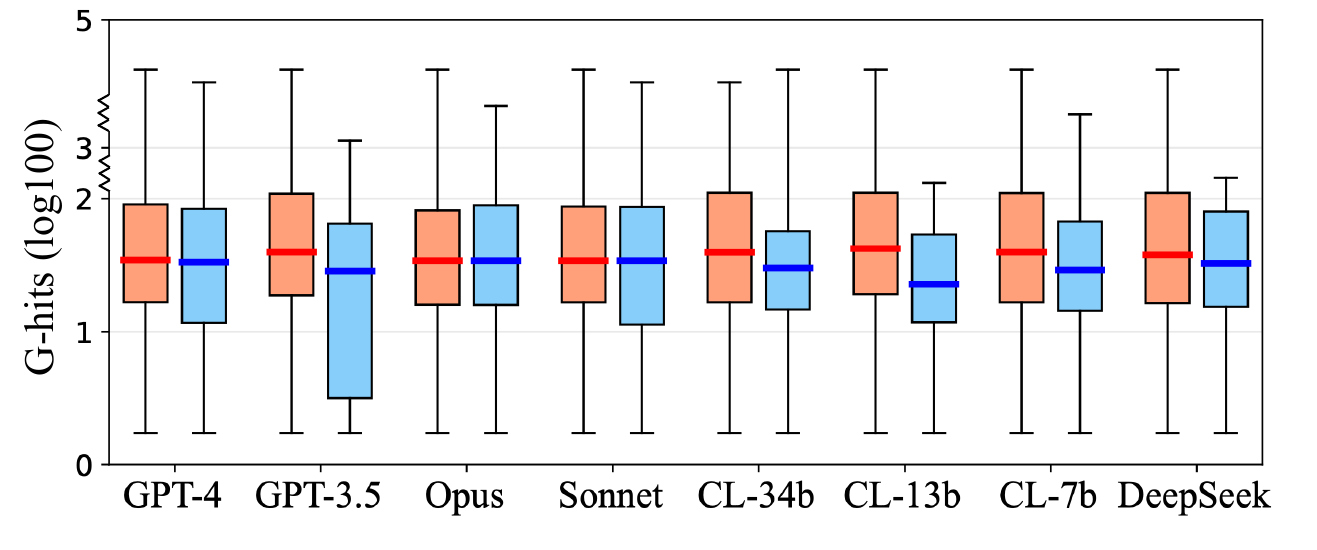}
    \vspace{-15pt}
    \caption{The \ghit distribution of the correctly detected misconfigurations (orange), and the \ghit distribution of the missed misconfigurations (blue).
    The bars in box plots indicate medians. CL refers to CodeLlama.}
    \label{fig:Ghits Wrong}
    \vspace{10pt}
\end{figure}

We study the correlation between a parameter's \ghit 
    and the effectiveness of LLMs in detecting the misconfigurations. 
For each configuration file in the \misconf dataset,
    we track the frequency of \llms detecting the parameter's misconfigurations 
    with the $i^{th}$ highest \ghit in each file, where $i=1...8$.
We separate cases when the misconfigured parameter is detected versus missed. 
As shown in Figure~\ref{fig:Ghits Wrong}, the median \ghit of misconfigured parameters being detected 
    is higher than the median \ghit of misconfigured parameters being missed.

\begin{figure}[t!]
   \centering
   \includegraphics[width=\linewidth]{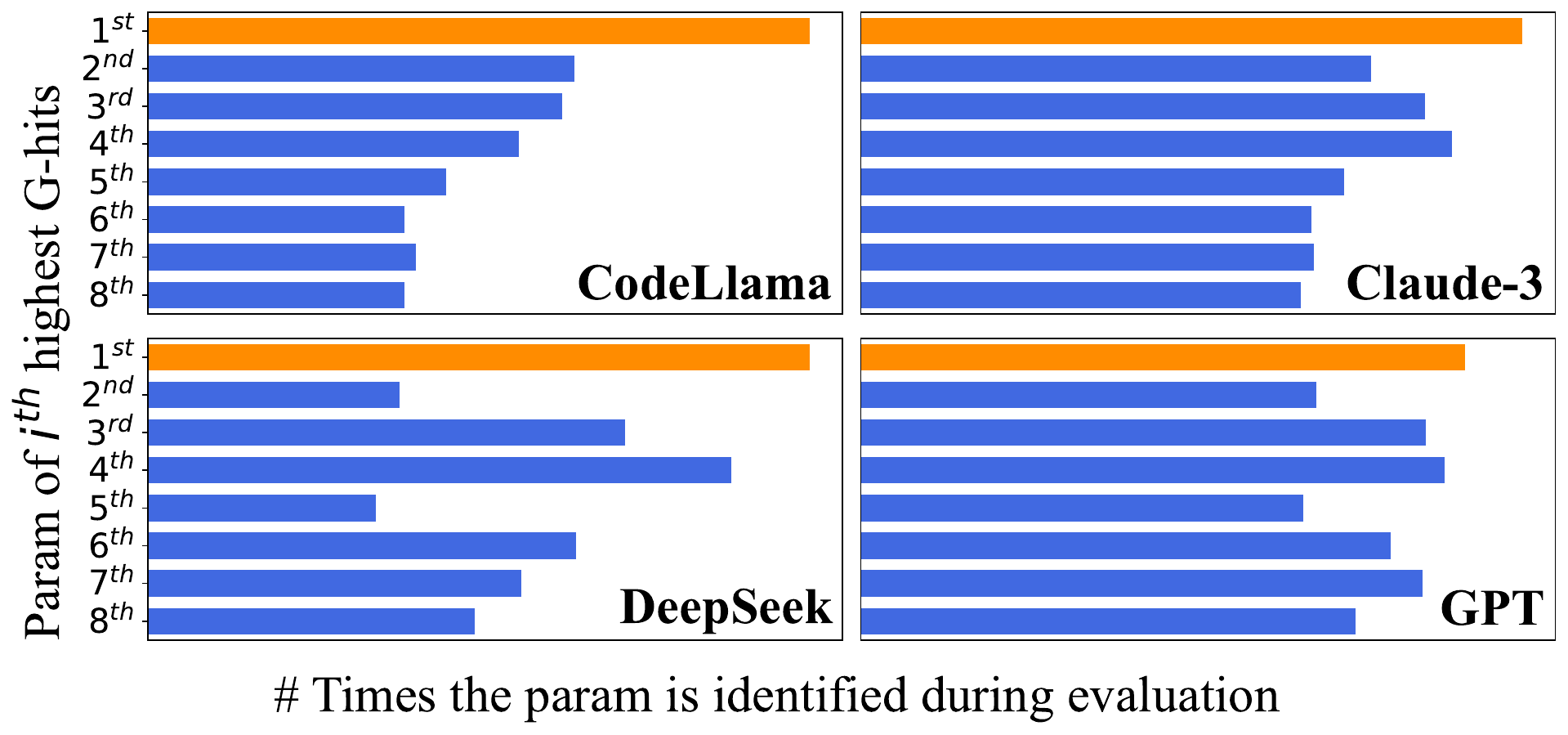}
   \vspace{-15pt}
   \caption{Frequency of the identified parameter with $i^{th}$ highest G-hits in a configuration file.}
   \label{fig:Ghits Correct}
   \vspace{5pt}
\end{figure}

We also study the frequency of false alarms
    across different ranking positions of \ghit within the file.
Specifically, for each configuration file in \validconfig dataset
    across all ten projects we evaluated, 
    we track the frequency of \llms mistakenly identifying 
    the parameter with the $i^{th}$ highest \ghit in each file,
    where $i=1...8$.
We group the results by the model family as shown in Figure~\ref{fig:Ghits Correct}.
The distributions reveal a clear skewness towards parameters with higher \ghit,
    indicating that \llms are more prone to report false alarms on popular parameters.

The biases can be attributed to the training data of \llms, which are from public
    domains 
    easily accessible by search engines like Google. 
Topics or parameters that are popularly discussed are more likely to be memorized by the \llms, 
    due to more frequent presence in the training data.
So, for configuration validation, 
    \llms can be less effective for parameters that are not commonly referenced online.

\section{Threats To Validity}\label{sec:threats}

\para{External threats.}
We evaluate \fw with eight state-of-the-art \llms to mitigate threats on evaluated models.
To mitigate threats of evaluated projects, 
    we select ten mature, widely used projects 
    of different types.
These systems are commonly used in prior studies~\cite{wang:icse:23,sun:osdi:20,cheng:issta:21,zhang:icse:21,rabkin:11,rabkin2:11,chen:fse:20}.
To account for bias in the evaluated configuration data, 
    we include many types of configuration parameters 
    and their generation rules based on prior work~\cite{xu:sosp:13,li:18,keller:dsn:08,li:issta:21}.
Our results cannot generalize to environment-related misconfigurations (discussed in \S\ref{sec:discussion}).
We expect the overall trend to be general, but the precise numbers may vary with 
    other \llms, projects, and configuration data in the field.

\para{Internal threats.} 
The internal threats lie in potential bugs in the implementation of \ourframework, and experimental scripts for evaluation. 
We have rigorously reviewed our code and multiple authors cross-validated the experiment results. 

\para{Construct threats.}
The threats to construct validity mainly lie in metrics (\S\ref{sec:meth:metrics}). 
We use the popular \fonescore, precision, and recall, 
and define our confusion matrices at both configuration file 
    and parameter levels.

\section{Discussion and Future Work}
\label{sec:discussion}

\para{Improving effectiveness of LLMs as validators.}
Despite the promising results, using LLMs directly as configuration validators like \fw 
    is a starting point to harness the ability of LLMs for configuration validation.
Specifically, there are circumstances where LLMs
    show limitations and biases (\S\ref{sec:eval:diff}, \S\ref{sec:eval:bias}).
One intricate aspect of configuration validation 
    is understanding \config dependencies. 
Integrating LLMs with configuration dependency 
    analysis~\cite{chen:fse:20} could be beneficial. 

We plan to investigate
    advanced prompting techniques, such as Chain-of-Thoughts (CoT)~\cite{wei2022chain,wang:arxiv:23,zhang2022automatic}.
For \config validation, CoT prompting can potentially mimic the reasoning process of
    a human expert.
By eliciting LLMs to generate intermediate reasoning steps toward the validation results, 
    it makes the validation more transparent and potentially more accurate. 
We also plan to explore extending Ciri into a multi-agent framework, 
    where Ciri can interact with additional tools such as 
    Ctest~\cite{sun:osdi:20} and Cdep~\cite{chen:fse:20} through agent frameworks such as LangChain~\cite{langchain} and AutoGen~\cite{wu2023autogen}.

    
Lastly, integrating user feedback loops can be valuable. 
With user feedback on validation results, 
    the iterative procedure can refine LLMs over time, 
    leading to more accurate responses.


\para{Detecting environment-related misconfigurations.}
While our study primarily targets misconfigurations that are common in the field~\cite{yin:sosp:11}, 
    the validity of configuration files can vary across environments.
For instance, a configuration parameter can specify a file path, 
    so the file's existence, permission, and content decide its validity.
To address these, \llms can generate environment-specific scripts to run in the target environment.
For example, given the configuration file as input, 
    the \llm can generate a Python script as follows. 


\begin{figure}[h!]
    \centering
        \includegraphics[width=0.85\linewidth]{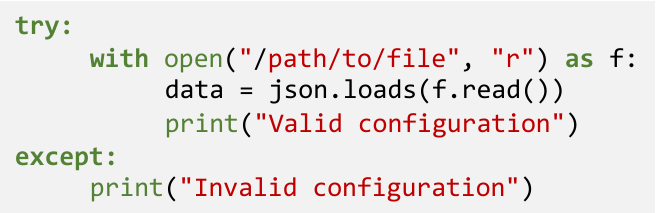}
\end{figure}    

Such LLM-generated scripts can help identify issues like misconfigured paths, unreachable addresses, 
    missing packages, or invalid permissions. Notably, these scripts offer a lightweight alternative 
    to configuration tests~\cite{xu:19,cheng:issta:21}.


\para{Detecting source-code related misconfigurations.}
We explored augmenting LLMs 
    with code snippets (Finding~\ref{finding:code-retrieval}),
    which can reveal parameter types and semantics.
This approach can be further improved by integrating advanced 
    program analysis
to present both configuration and relevant source code to the \llm. 
Techniques like static or dynamic program slicing~\cite{attariyan:osdi:10,zhang:oopsla:21,xu:osdi:16,rabkin2:11} 
can help identify the relevant code. 

\para{Fine-tuning LLMs for configuration validation.} 
We also plan to explore fine-tuning to tackle system-specific configuration problems,
    which is hard to address with common-sense knowledge. 
Specifically, configuration related software evolution is prevalent, which 
    introduces new parameters and changes the semantics and constraints of existing parameters~\cite{zhang:icse:14,zhang:icse:21}.
A promising solution is to fine-tune LLMs on new code/data, and make LLMs evolution-aware.

\section{Related Work}
\label{sec:related-work}

%

\para{Configuration validation.}
Prior studies developed frameworks for developers to implement validators~\cite{tang:sosp:15,raab:17,baset:middleware:2017,huang:15}
    and test cases~\cite{sun:osdi:20,xu:19},
    as well as techniques to extract configuration constraints~\cite{liao:18,NadiTSE2015,xu:sosp:13,zhang:oopsla:21}.
However, manually writing validators and tests requires extensive engineering efforts, 
    and is hard to cover various properties of different configurations~\cite{xu:sosp:13,xu:osdi:16,li:18,keller:dsn:08,li:issta:21}.
ML/NLP-based configuration validation techniques have been investigated.
Traditional ML/NLP-based approaches learn correctness rules from configuration data~\cite{Bhagwan:21,zhang:asplos:14,wang:osdi:04,le:06,palatin:kdd:06,wang:lisa:03,santo:16,santo:17} 
    and documents~\cite{xiang:atc:20,potharaju:vldb:15} and then use the learned rules for validation. 
These techniques face data challenges and rely on predefined learning features and models, 
    making them hard to generalize to different projects and deployment scenarios.
We explore
    using LLMs for configuration validation, which can potentially address the limitations 
    of traditional ML/NLP techniques towards automatic, effective validation solutions.
    

\para{LLMs for software engineering.}
\llms are 
actively applied to software engineering tasks, 
where they have demonstrated effectiveness in generating, summarizing, and translating code~\cite{chen:arxiv:21,li:22,lyer:arxiv:18,
    ahmed:ase:22,lu:arxiv:21,roziere:nips:20,roziere:arxiv:21}, 
    failure diagnosis~\cite{chen:arxiv:23,ahmed:icse:23}, 
    fault localization and program repair~\cite{mirsky:ss:21,fan:icse:23,xia:icse:23,xia2023universal}.
LLMs for code are also increasingly prominent~\cite{feng:arxiv:20,chen:arxiv:21,nijkamp:arxiv:20,fried:arxiv:23,xu:arxiv:22,nedelkoski2019anomaly}, 
    and are used for coding tasks.
We take a first step to apply LLMs for software configuration problems,
    and show that LLMs have the potential to efficiently automate certain validation tasks
    and even bring advances over developer-written validators.
Ciri as a framework is generic to different LLMs. 

\section{Concluding Remarks}

As a first step to harvest LLMs for software configuration,
    we develop \fw as an open platform to experiment with LLMs as configuration validators,
    and present the important design choices.
Through Ciri, 
    we analyze LLM-empowered configuration validators. 
Our analysis shows the promising effectiveness of state-of-the-art LLMs as configuration validators,
    as well as their limitations.
Our work shed light on new, exciting research directions of using 
    LLMs for software configuration research.

\bibliographystyle{acm}
\bibliography{ref} 

\end{document}